\documentclass[10pt, conference]{IEEEtran}
\IEEEoverridecommandlockouts
\usepackage{cite}
\usepackage{amsmath,amssymb,amsfonts}
\usepackage{graphicx,setspace}
\usepackage{textcomp}
\usepackage{xcolor}
\usepackage{algorithm,algpseudocode}
\newcommand\CONDITION[2]%
  {\begin{tabular}[t]{@{}l@{}l@{}}
     #1&#2
   \end{tabular}%
  }
\algdef{SE}[WHILE]{While}{EndWhile}[1]%
  {\algorithmicwhile\ \CONDITION{#1}{\ \algorithmicdo}}%
  {\algorithmicend\ \algorithmicwhile}
\algdef{SE}[FOR]{For}{EndFor}[1]%
  {\algorithmicfor\ \CONDITION{#1}{\ \algorithmicdo}}%
  {\algorithmicend\ \algorithmicfor}
\algdef{S}[FOR]{ForAll}[1]%
  {\algorithmicforall\ \CONDITION{#1}{\ \algorithmicdo}}
\algdef{SE}[REPEAT]{Repeat}{Until}{\algorithmicrepeat}[1]%
  {\algorithmicuntil\ \CONDITION{#1}{}}
\algdef{SE}[IF]{If}{EndIf}[1]%
  {\algorithmicif\ \CONDITION{#1}{\ \algorithmicthen}}%
  {\algorithmicend\ \algorithmicif}%
\algdef{C}[IF]{IF}{ElsIf}[1]%
  {\algorithmicelse\ \algorithmicif\ \CONDITION{#1}{\ \algorithmicthen}}
\usepackage{textcomp}
\usepackage{array}
\usepackage{tabularx}
\usepackage{booktabs}
\usepackage{mathtools}
\usepackage{comment}
\usepackage{amsthm}
\usepackage{url}
\usepackage{siunitx}

\usepackage{epstopdf}

\usepackage{subfigure}
\DeclareMathOperator{\dist}{dist} 
\DeclarePairedDelimiter{\norm}{\lVert}{\rVert}

\allowdisplaybreaks
\begin{document}
\title{Disconnectivity-Aware Energy-Efficient Cargo-UAV Trajectory Planning with Minimum Handoffs
\thanks{This 
work is funded in part by Huawei Canada and in part by the Natural 
Sciences and Engineering Council Canada (NSERC).}}
\author{\IEEEauthorblockN{  Nesrine~Cherif\IEEEauthorrefmark{1}, Wael Jaafar\IEEEauthorrefmark{2}, Halim Yanikomeroglu\IEEEauthorrefmark{2}, and Abbas Yongacoglu\IEEEauthorrefmark{1}. \\
	\IEEEauthorblockA{\IEEEauthorrefmark{1}School of Electrical Engineering and Computer Science, University of Ottawa, Ottawa, ON, Canada.\\
		\IEEEauthorrefmark{2}Department of Systems and Computer Engineering, Carleton University, Ottawa, ON, Canada.	
		}
		}
}

\maketitle
\linespread{1.1}
\begin{abstract}
On-board battery consumption, cellular disconnectivity, and frequent handoff are key challenges for unmanned aerial vehicle (UAV) based delivery missions, a.k.a., cargo-UAV. Indeed, with the introduction of UAV technology into cargo shipping and logistics, 
designing energy-efficient paths becomes a serious issue for the next retail industry transformation. Typically, the latter has to guarantee uninterrupted or slightly interrupted cellular connectivity for the UAV's command and control through a small number of handoffs. 
In this paper, we formulate the trajectory planning as a multi-objective problem aiming to minimize both the UAV's energy consumption and the handoff rate, constrained by the UAV battery size and disconnectivity rate. 
Due to the problem's complexity, we propose a dynamic programming based solution.
Through simulations, we demonstrate the efficiency of our approach in providing optimized UAV trajectories. Also, the impact of several parameters, such as the cargo-UAV altitude, disconnectivity rate, and type of environment, are investigated. The obtained results allow to draw recommendations and guidelines for cargo-UAV operations. 

\end{abstract}

\section{Introduction}
Cargo unmanned aerial vehicle (cargo-UAV) is a popular mean of delivery, especially for remote and hard-to-reach areas \cite{amazonprimeair}. Due to the technological advancements that made cargo-UAVs safer and cost-efficient, it is expected that the latter be deployed in dense urban centres \cite{cherifhighway2020}. 
Since cargo-UAVs operate at beyond visual line-of-sight (BVLoS), command and control (C\&C) is crucial in ensuring the safety and security of the mission \cite{fotouhiSurvey}. 
The latter relies on cellular networks, which have to be reliable enough for safe operation.
Several works have investigated the practicality and feasibility of BVLoS cellular-connected UAVs \cite{Ozger2018,Li2019,Amorim2020}.

Reusing the terrestrial networks for cargo-UAV cellular connectivity presents its own set of challenges as these networks were primarily designed to serve ground users only \cite{cherifhighway2020}. Although economically attractive, it may be difficult to provide reliable and seamless cellular connectivity to the cargo-UAVs, especially at high altitudes.
Indeed, terrestrial base stations (BSs) are down-tilted, which means that cargo-UAVs can be served only through the sidelobes of BS antennas \cite[Fig. 6]{cherifdownlink}. 
Moreover, the cargo-UAV's inherited mobility leads to frequent switching from one serving BS to another, i.e., handoff. As such, an additional amount of information is exchanged, which degrades further the transmission link. 
Even though the energy consumption related to the communication overhead is relatively small compared to the cargo-UAV's motion energy, handoff events should be kept to the minimum in order to guarantee the C\&C stability.
Due to the aforementioned factors, occasional cellular disconnectivity can be experienced by cargo-UAVs operating in BVLoS. Depending on the criticality of the mission and the level of autonomy, a cargo-UAV can tolerate an end-to-end disconnectivity rate, defined as the ratio of the time spent disconnected from all cellular BSs to the total mission time.
\subsection{Related Work}

In order to keep low handoff and disconnectivity rates, cargo-UAV trajectory planning became significantly important in BVLoS missions. 
This problem has attracted many researchers for the last few years.
For instance, the authors of \cite{fakhreddine2019handover} set up a testbed for cellular-connected UAV missions in the sky. Their qualitative study showed that handoff occurs more frequently for UAVs, compared to terrestrial users. Hence, proper UAV motion optimization is needed to reduce handoffs. A cellular disconnectivity-aware trajectory design was proposed in \cite{bulutdisc2018}. However, the authors relied rather on a simplistic cellular-to-air (C2A) propagation model that does not reflect the real 3D radiation patterns of BSs. 
In contrast, the authors of \cite{chowdhury20203d} studied the effects of both realistic 3D antenna radiation and limited backhaul on the UAV trajectory planning problem. Yet, this work lacked the insights on the handoff and disconnectivity rates during UAV missions. Recently, reinforcement-learning based trajectory planning solutions have been proposed for  energy-efficiency maximization \cite{challita2018drl}, mission time minimization \cite{zengpath2019}, and route distance minimization \cite{zhang2019radio}. However, none of these works
has accounted for the UAV energy consumption, cellular disconnectivity, and handoffs, altogether. This has been investigated for the first time in \cite{azari2020mobile}, where
a multi-armed-bandit based UAV velocity strategy is proposed for a predefined trajectory, and under a weighted handoff rate, cellular disconnectivity, and power consumption objective. However, the trajectory was not optimized with  consideration of these factors.  
\subsection{Contributions}

Given the limitations of the aforementioned works, we study in this paper the disconnectivity and handoff aware trajectory planning problem for cargo-UAVs. 
Specifically, we aim to minimize the UAV's energy consumption and handoff rate, with respect to the battery size and disconnectivity rate. Due to the problem's complexity, we propose a dynamic programming (DP) based solution that provides reliable cargo-UAV trajectories. Through simulations, the efficiency of our method is demonstrated. Finally, by evaluating the impact of several parameters, e.g., cellular connectivity conditions and UAV altitude, we provide design guidelines for cargo-UAV operations.   

The remaining of the paper is organized as follows. Section II presents the system model.
Section III formulates the trajectory planning problem. Section IV presents the proposed DP-based solution, while the simulation results are discussed in Section V. Finally, Section VI closes the paper.

\section{System Model}
\label{sec:sysmodel}
We assume a geographical area with its airspace, defined through a 3-dimensional (3D) axes system $(x,y,z)$. As per the 3GPP report \cite{3gpp777}, a cargo-UAV is allowed to hover/fly at a maximum altitude of 300 m above the ground level. We assume that  the cargo-UAV mission is to deliver an item from a retailer's warehouse, located in $(x_s,y_s)$ in the $(x,y)$ plane to a consumer's drop-off location $(x_d,y_d)$. We envision that the cargo-UAV hovers vertically from the warehouse's dockstation until it reaches the desired altitude $h \leq$ 300 m at coordinates $\text{START}=(x_s,y_s,h)$, then it travels horizontally with a fixed velocity $v$ to reach the destination point, defined as $\text{END}=(x_d,y_d,h)$. Finally, the cargo-UAV lands vertically to the drop-off location $(x_d,y_d)$. 

We assume that the cargo-UAV is exclusively served by the terrestrial cellular network. The latter is constituted of
$M$ BSs, where the location of the $i^{th}$ BS is denoted by $(x_{\rm BS}^i,y_{\rm BS}^i,z_{\rm BS})$. The height of the BS depends on the type of environment, which is selected from 35 m, 25 m, and 10 m, for rural macro (RMa), urban macro (UMa), and urban micro (UMi), respectively. 
\vspace{-0.25cm}
\subsection{C2A Channel Model}
The C2A channel model is outlined in the 3GPP reports \cite{3gpp777,3gpp36814}. It considers several parameters, including the BS's antenna gain, LoS probability, and path-loss.

\subsubsection{BS Antenna Gain}
As mentioned previously, UAVs in the sky are mainly connected to BSs through the radiating sidelobes. Hence, 
accurately modelling the BS 3D antenna radiation pattern is needed for accurate trajectory planning. In this work, we adopt the 3GPP antenna pattern model that mimics realistic antenna patterns \cite{3gpp36814}. That means, each BS is divided into three sectors, where a sector is equipped with cross-polarized antennas to form a uniform linear array (ULA). Each antenna element provides a gain up to $G_{\max}$ (typically 8 dBi) in the direction of the mainlobe.
The antenna element pattern is given by \cite[Table 7.1.-1]{3gpp36814}, which provides certain gains depending on the azimuth and elevation angles of the cargo-UAV with respect to the BS's location. They are respectively given by  
\vspace{-0.2cm}
\begin{equation}
G_{\rm az}(\phi_i)=\min\left\{12\left(\frac{\phi_i}{\phi_{\rm 3dB}}\right), G_m\right\},
\end{equation}
\begin{equation}
G_{\rm el}(\theta_i)=\min\left\{12\left(\frac{\theta_i}{\theta_{\rm 3dB}}\right), {\rm SLA}\right\},
\end{equation}
where $\phi_i=\tan^{-1}\left(\frac{h-z_{\rm BS}}{d_{2D,i}}\right)$ and $\theta_i=\tan^{-1}\left(\frac{y-y_{\rm BS}^i}{x-x_{\rm BS}^i}\right)$ are the azimuth and elevation angles between the $i^{th}$ BS and the cargo-UAV. Also, $d_{2D,i}$ is the horizontal distance between the $i^{th}$ BS and the cargo-UAV, while $\phi_{\rm 3dB}$ and $\theta_{\rm 3dB}$ are the 3 dB bandwidths fixed at $65^{\circ}$. Furthermore, $G_m$ and ${\rm SLA}$ are the thresholds for antenna nulls, fixed to 30 dB. Finally, the antenna element gain in dB is expressed by 
\begin{equation}
    G(\theta_i,\phi_i)=G_{\max}-\min \left\{-\left(G_{\rm az}(\phi_i)+G_{\rm el}(\theta_i) \right), G_m\right\}.
\end{equation}

Assuming $N$ antenna elements at BS $i$, equally distant by half of the wavelength, the array factor ${\rm AF}$ of the ULA with a down-tilt of $\theta_d$ can be given by \cite[eq. (2)]{xu2019icc}
\begin{equation}
    \label{eq:af}
    {\rm AF} (\theta_i)=\frac{\sin\left(\frac{N\pi}{2}\left(\sin{\theta_i}-\sin{\theta_d}\right)\right)}{\sqrt{N}\sin\left(\frac{\pi}{2}\left(\sin{\theta_i}-\sin{\theta_d}\right)\right)},\; \forall i=1,\ldots,M.
\end{equation}
Finally, the array radiation pattern of the $i^{th}$ BS   is written by
\vspace{-0.2cm}
\begin{equation}
    \label{eq:ar}
    G_i=G(\theta_i,\phi_i)+{\rm AF}(\theta_i),\; \forall i=1,\ldots,M.
\end{equation}

\subsubsection{LoS Probability}
The probability that the  cargo-UAV is in LoS condition with the BS depends mainly on the terrestrial environment and the cargo-UAV altitude.
For instance, when 22.5 m $< h \leqslant$ 100 m, the probability of LoS, denoted $P_{\rm LoS}(d_i)$ in UMa is given by \cite[Table B-1]{3gpp777}
\begin{equation}
\label{eq:probLoS}
P_{\rm LoS}(d_i) =\left\{
\begin{array}{ll}
\!\!\!\! 1 ,& d_{2D,i}\leqslant d_1\\
\!\!\!\!\frac{d_1}{d_{2D,i}}\!+\!\exp\left(\!\frac{-d_{2D,i}}{p_1}\!\right) \left(\!1\!-\!\frac{d_1}{d_{2D,i}}\!\right), & d_{2D,i} > d_1,
\end{array}
\right.
\end{equation}
where $d_i=\sqrt{d_{2D,i}^2+(h-z_{\rm BS})^2}$ is the 3D distance between the cargo-UAV and the $i^{th}$ BS, $p_1=4300 \log_{10}(h)-3800$, and $d_1=\max \left\{460 \log_{10}(h)-700,18\right\}$. When 100 m $<h\leq$ 300 m, $P_{\rm LoS}(d_i)=1$. Finally, the NLoS probability is calculated as $P_{\rm NLoS}(d_i)=1-P_{\rm LoS}(d_i)$.

\subsubsection{Pathloss Model}
For the sake of simplicity, we follow the probabilistic pathloss model, as defined in
\cite[Tables B-1 and B-2]{3gpp777}. The average pathloss of the channel between the cargo-UAV and the $i^{th}$ BS, denoted $L_i$, is given by  
\begin{equation}
    \label{eq:pathloss}
    {L_i}=P_{\rm LoS}(d_i) L_i^{\rm LoS} +P_{\rm NLoS}(d_i) L_i^{\rm NLoS},\; \forall i=1,\ldots,M,
\end{equation}
where $L_i^{\rm LoS}$ and $L_i^{\rm NLoS}$ are the pathlosses related to fully LoS and NLoS communication links, respectively, defined in \cite[Tables B-1 and B-2]{3gpp777}.

\subsection{Received Power Analysis}

Based on the C2A channel model and assuming that the cargo-UAV is equipped with an omni-directional antenna with unit gain, the average received power from the $i^{th}$ BS is given by \vspace{-0.3cm}
\begin{equation}
    \label{eq:Pr}
    P_{r,i}=P_T + G_i-L_i, \; \forall i=1,\ldots,M,
\end{equation}
where $P_T$ is the transmit power of any BS in dBm. 

The received power from a BS at any 3D location determines the quality of the transmission link. As discussed in \cite{3gpp777}, the airspace is an interference-limited environment since a cargo-UAV may have cleared LoS conditions with several BSs using the same frequency resource at the same time\footnote{In this paper, we investigate the full interference scenario. Nevertheless, frequency planning can be considered to reduce the number of inter-cell interferers.}.
Thus, we assume for simplicity that the cellular connectivity depends solely on the level of the signal-to-interference ratio ($\textsf{SIR}$), defined by
\vspace{-0.4cm}

\begin{equation}
    \textsf{SIR}_i=\frac{P_{r,i}}{\sum \limits_{\substack{j=1,j \neq i}}^{M} P_{r,j}},\; \forall i=1,\ldots,M.
\end{equation}
One location is presumed a \textit{coverage hole} if $\textsf{SIR}$ falls below a predefined threshold $\beta_{\rm th}$.
\begin{figure}[t]
	\centering
	\includegraphics[width=0.98\linewidth]{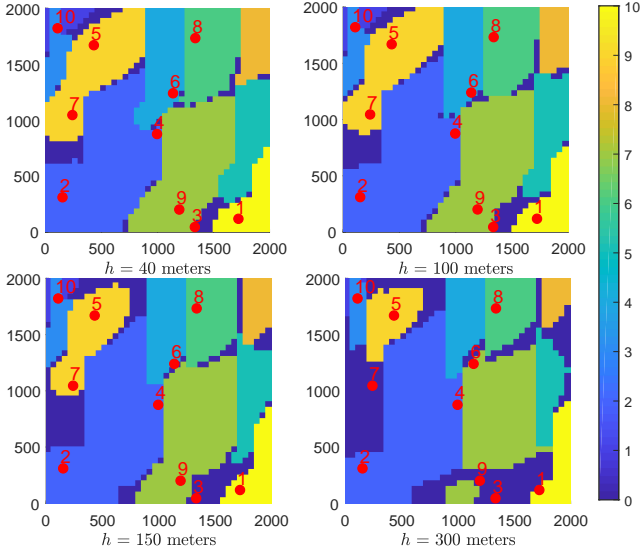}
	\caption{Cellular connectivity heatmap for different cargo-UAV altitudes. Colors denote the IDs of the serving BSs. ID 0 (dark blue) defines the coverage holes ($\theta_{d}= -10^{\circ}$, $f_c=2$ GHz, $v=30$ m/s, and $\beta_{\rm th}=0$ dB).}
	\label{fig:heatmap}
\end{figure}

Using (\ref{eq:Pr}), we depict in Fig. \ref{fig:heatmap} the 3D cellular coverage of 10 BSs at different cargo-UAV altitudes. We assume here an area of $2 \times 2$ km$^2$, and the coverage at any 3D location is provided by the BS with the strongest $\textsf{SIR}$, i.e., $\textsf{SIR}=\max_{i=1,\ldots,M}\textsf{SIR}_i$. 
We notice that the coverage area of the BS $10$ at the north-east corner (clear yellow color) is located at the south-east corner of the map. This observation corroborates the fact that BSs serve cargo-UAVs using the sidelobes.
Moreover, the coverage holes widen while the coverage areas shrink as the cargo-UAV flies higher. This is due to the large pathloss and low antenna gain at high elevation angles.

\subsection{Cargo-UAV Energy Model}
The energy consumed by a cargo-UAV depends on its size, weight, and power (SWAP), in addition to its motion and communication regimes \cite{jaafar2020dynamics}. It is composed of the propulsion energy, needed for flying, and of the communication energy, for C\&C operations. 
Typically, communication energy is very small compared to the propulsion energy, and thus, can be neglected. Using a rotary-wing cargo-UAV, the propulsion power can be given by \cite{zeng2019}
\vspace{-0.25cm}
\begin{eqnarray}
\label{eq:prop}
P_{\rm prop}(v)&=&\frac{1}{2}\delta_0 \zeta r A v^3+\sigma_B \left(1+\frac{3v^2}{U_{\rm tip}^2}\right)\nonumber \\
&+& \sigma_I \left({\sqrt{1+\frac{v^4}{4v_0^4}}-\frac{v^2}{2v_0^2}}\right)^{\frac{1}{2}}, 
\end{eqnarray}
where $\sigma_B$ and $\sigma_I$ are the blade profile power and induced power, respectively. $v_0$ is the mean rotor induced velocity and $U_{\rm tip}$ is the speed of the tip of the rotor blade. Also, $\zeta$ is the air density, $\delta_0$ is the fuselage drag ratio, $r$ is the rotor solidity, and $A$ is the rotor disc area. Subsequently, the propulsion energy to fly for a distance $d$ is
\vspace{-0.3cm}
\begin{equation}
    \label{eq:ener}
    E(v,d)={P_{\rm prop}}(v) \; d/{v}, \; v>0.
\end{equation}
We denote by $E_C$ and $E_S$
the initial cargo-UAV energy capacity and emergency reserve energy, respectively.
$E_C$ is available for the mission, while $E_S$ is used in emergency pull-backs. Hence, the cargo-UAV available energy to travel between the START and END locations is
\vspace{-0.3cm}
\begin{equation}
    E_A(v,h)=E_C-E_S-2 E(v,h),
\end{equation}
where $E(v,h)$ is the required energy for tackoff/landing up to/down to the dockstation/drop-off location. Thus, the maximum horizontal travelling distance is given by 
\vspace{-0.25cm}
\begin{equation}
\label{eq:maxDist}
d_{\max}= {v E_A(v,h)}/{P_{\rm prop}(v)}.
\end{equation}

\section{Problem Formulation}
The cargo-UAV mission is to deliver an item from a warehouse to a consumer drop-off location. Assuming that the cargo-UAV flies at a constant speed and altitude $h$, we aim to design its mission trajectory such that: 1) it consumes the lowest energy, 2) it minimizes the handoff rate, 3) it does not exhibit a cellular disconnectivity rate above a fixed threshold $\alpha_{\max}$ during the mission. 
Based on the cellular connectivity heatmap shown in Fig. \ref{fig:heatmap}, a disconnectivity event happens when the cargo-UAV transits to/from an area with BS ID 0 (coverage hole), and the \textit{disconnectivity rate} is calculated as the ratio of the time spent in areas with BS ID 0 to the total mission time.
When the cargo-UAV transits from a coverage area to another (with BS IDs $\neq$ 0), a handoff event occurs. Then, the \textit{handoff rate} is calculated as the total number of handoff events during the mission. Let $\mathcal{T}=\{\textbf{q}_1,\ldots, \textbf{q}_{|\mathcal{T}|}\}$ be the trajectory of the cargo-UAV, defined by the locations $\textbf{q}_i=(x_i,y_i,h)$. Also, let $\mathcal{S}=\{s_1,\ldots, s_{|\mathcal{T}|}\}$ be the set of BS IDs associated to the cargo-UAV locations in $\mathcal{T}$. Subsequently, the trajectory planning problem can be formulated as follows:
\vspace{-0.4cm}
\begin{subequations}
	\begin{align}
	\min_{\mathcal{T}} \quad & w_1 \sum_{i=1}^{|\mathcal{T}|-1} E(v,d_{i,i+1})+ w_2\sum_{i=1}^{|\mathcal{T}|-1}\eta_{i,i+1}\tag{P1} 
	\\
	\label{c2_21} \text{s.t.}\quad & \sum_{i=1}^{|\mathcal{T}|-1} E(v,d_{i,i+1}) \leq E_{A}(v,h),  \nonumber \tag{P1.a}\\
	\label{c2_2} &\frac{d_{\rm disc}}{\sum \limits_{i=1}^{|\mathcal{T}|-1} d_{i,i+1}} \leq \alpha_{\max}, \tag{P1.b}
	\end{align}
\end{subequations}
where $w_1$ and $w_2$ are weight factors balancing between energy consumption and handoff rate. $d_{i,i+1}=\norm{\textbf{q}_{i}-\textbf{q}_{i+1}}$, $d_{\rm disc}$ is the sum of flying distances in coverage holes, and $\eta_{i,i+1}$ is a binary variable reflecting the handoff event, i.e., $\eta_{i,i+1}=1$ if $s_i \neq s_{i+1}$ and $(s_i,s_{i+1}) \neq (0,0)$, and $\eta_{i,i+1}=0$ otherwise.
Problem (P1) is NP-hard. Indeed, the latter can be seen as the capacitated vehicle routing problem (CVRP) \cite{Lenstra1981}. The CVRP is described as selecting the route for a vehicle, aiming to serve a group of customers. The vehicle has a limited capacity, which is used to depart from a depot point, serve a number of customers along its route, then return to the same or a different depot point. The objective of the CVRP is to minimize the total transport costs. Logically, the vehicle, customers, and transport costs, can be assimilated by the UAV, BSs, and (energy consumption + handoff rate), respectively. Since the CVRP is known to be NP-hard \cite{Lenstra1981}, then by restriction, (P1) is also NP-hard.

\section{Proposed Solution: DP Algorithm}
In this section, we adopt, similarly to \cite{bulutdisc2018}, the dynamic programming approach to derive a near-optimal trajectory solution for (P1).

\begin{figure}[t!]
	\centering
	\includegraphics[width=0.2\linewidth]{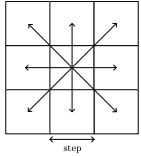}
	\caption{Motion options for the cargo-UAV in one step.}
	\label{fig:grid}
\end{figure}

First, we start by dividing the plane $(x,y)$ at altitude $h$ to a grid of cells, numbered from 1 to $J$. We assume that the cargo-UAV can move in one step from the center of one cell to the center of an adjacent cell in one of the available directions, i.e., one of 8 movements can be realized as shown in Fig. \ref{fig:grid}. We define by $\mathcal{T}_j$, $\mathcal{S}_j$, $d_{\rm disc}^j$, and $E_j$ the best trajectory to reach cell $j$ (with respect to energy and disconnectivity constraints), the corresponding set of BS IDs, the sum of flying distances in coverage holes to reach cell $j$ when following trajectory $\mathcal{T}_j$, and the associated cargo-UAV consumed energy, respectively. Thus, for any cell $j$, the objective function can be computed as
\vspace{-0.4cm}
\begin{eqnarray}
    \label{eq:D_j}
    O_j(t)=\min_{j' \in \mathcal{N}_j} \left\{ O_{j'}(t-1)+w_1 E(v,d_{j',j})+w_2 \eta_{j',j} \right\}, \\  \text{such that } E_{j'}+E(v,d_{j',j})\leq E_A; \; \frac{d_{\rm disc}^{j}}{\sum \limits_{i=1}^{|\mathcal{T}_j|-1}d_{i,i+1}}\leq \alpha_{\max}, \nonumber
\end{eqnarray}
where $t$ is the stage in the optimization process, $\mathcal{N}_j$ is the set of neighbouring cells to cell $j$, $\eta_{j',j}$ is the binary variable indicating a handoff event, and
\vspace{-0.3cm}
\begin{equation}
 O_j(t)=w_1 \sum_{i=1}^{|\mathcal{T}_j|-1} E(v,d_{i,i+1})+w_2 \sum_{i=1}^{|\mathcal{T}_j|-1} \eta_{i,i+1},\; \forall j.   
\end{equation}
\vspace{-0.2cm}

Eq. (\ref{eq:D_j}) states that if the cargo-UAV can reach the $j^{th}$ cell from a neighboring cell while 1) minimizing the objective function, 2) not exceeding the available on-board energy, and 3) not exceeding the disconnectivity rate threshold, then, the new optimal trajectory is from that neighboring cell. When 
repeated with a sufficient number of iterations, this
step provides the optimal trajectory.  

In Algorithm \ref{algo:dp}, we present the pseudocode for the DP-based trajectory planning method. In lines 2 and 3, the minimum cost path matrices are initialized and computed. Then, the stage cost matrices are initialized in lines 4 and 5, where the initial cost of the START cell is set to zero. In line 6, we generate the matrix in which we will store the optimal cells indices guaranteeing the smallest costs. 
Lines 7 to 21 execute the DP algorithm iterations, where the battery, cellular disconnectivity, and handoff constraints are handled in line 13. Then, the solution is refined with a smaller cost at each iteration, as shown in line 14. Finally, the optimal path is extracted recursively in the procedure of lines 22 to 31.

\section{Simulation Results}
\begin{figure}[t]
\centering
\includegraphics[width=0.5\linewidth]{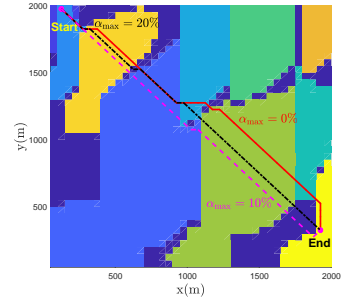}
\caption{Cargo-UAV paths ($h=200$ m, different $\alpha_{\max}$).}
\label{fig:path}
\end{figure}

In this section, we evaluate the performance of the cargo-UAV DP-based path planning solution. 
Unless stated elsewhere, we assume here 10 BSs deployed in an UMa environment  using the pathloss model of \cite[Table B-1]{3gpp777}. As shown in Fig. \ref{fig:path}, the START and END locations for the cargo-UAV mission are selected in the two opposite corners of the map. We set the following parameters to $G_{\max}=8$ dBi, $\theta_d=-10^{\circ}$, $P_T=43$ dBm, $v=\ 30$ m/s, $E_C-E_S=2500$ kJ, and $w_1=w_2=1$. Finally, typical values of the cargo-UAV related parameters are taken from \cite[Table I]{zeng2019}.

In Fig. \ref{fig:path}, we illustrate the resulting cargo-UAV paths using Algorithm \ref{algo:dp}, given different values of the tolerated disconnectivity $\alpha_{\rm max}$. The dark blue color refers to the coverage holes, while the other colors correspond to the coverage areas of 10 BSs, located as shown in Fig. \ref{fig:heatmap}. As $\alpha_{\max}$ decreases, we see that the followed path is longer. Indeed, the UAV needs to make detours in order to avoid partially ($\alpha_{\max}>0 \%$) or completely ($\alpha_{\max}=0 \%$) the coverage holes.

\begin{algorithm}[ht]
\caption{DP-based Trajectory Planning}
\label{algo:dp}
\small
\begin{algorithmic}[1]
\Require{Matrix $\mathcal{H}$ of size $(M\times N)$ representing cellular connectivity heatmap (e.g., Fig. \ref{fig:heatmap}), step size $\gamma$, $v$, $\alpha_{\max}$, $E_A$, $\max_{\rm iter}$, and START and END locations.}
\Ensure{Optimal trajectory $\mathcal{T}^*$.}
\State Compute $d_{\max}$ using (\ref{eq:maxDist}).
\State Initialize transition matrices $T_{\nu}$, $\nu \in \{\rm dist,disc,ho\}$ of size $(MN\times MN)$ with $\infty$.
\State Compute the cost (dist, disc, ho) from each cell in $\mathcal{H}$ to all other cells and store it in $T_{\nu}$, $\nu \in \{\rm dist,disc,ho\}$.
\State Initialize stage cost matrices ($S_{\dist},\ S_{\rm disc},\ \text{and }S_{\rm ho}$) of size $[1,M]$ with $\infty$.  
\State Set the initial cost  of the START cell to $0$, i.e., $S_{\nu}(\rm START)=0$, $\nu \in \{\rm dist,disc,ho\}$.
\State Generate the predecessor matrix $P_{\dist}$ of size $(M\times \max_{\rm iter})$ for recursion. \% $\max_{\rm iter}$ is the maximal number of DP iterations
  \For {$q=2$ to $\max_{\rm iter}$}
  \State{Store $S_{\nu}$ in predecessor matrix $S_{\nu}^{\rm pr}$, $\nu \in \{\rm dist,disc,ho\}$.}
  \For {$i = 1$ to $M$ }
     \For {$j = 1$ to $N$}
     \State  \parbox[t]{\dimexpr\linewidth-\algorithmicindent}{Compute the $\nu$ cost from cell $j$ to cell $i$,\\ i.e., $\nu_j=  T_{\nu}(j,i)+S_{\nu}^{\rm pr}(j)$, $\nu \in \{\rm dist,disc,ho\}$.}
     \If { the previous distance cost is higher than the \\computed one, i.e., $\rm{dist}_j\leqslant S_{\dist}(j)$}
     \If 
     {$\rm{dist}_j\leqslant d_{\max}$
     \& $\frac{{ {\rm disc}_j}}{{\rm dist}_{j}} \leqslant \alpha_{\max}$, \\ 
     \& ${{\rm ho}_j}\leqslant S_{\rm ho}(j)$}
     \State \parbox[t]{\dimexpr\linewidth-\algorithmicindent}{Update $S_{\nu}$ $S_{\nu}(j)=\nu_j$, $\nu \in \{\rm dist,disc,ho\}$. \strut}
     \State \parbox[t]{\dimexpr\linewidth-\algorithmicindent}{Store the optimal path from cell $j$ to cell $i$ \\in $P_{\dist}(j,q)=i$.\strut}
     \EndIf
     \EndIf
     \EndFor
     \EndFor
     \State \parbox[t]{\dimexpr\linewidth-\algorithmicindent}{Break if the END cell is reached.}
  \EndFor
  \State \parbox[t]{\dimexpr\linewidth-\algorithmicindent} {Set the current cell to END cell.}
  \State Set ${\rm st}=$ size($P_{\dist},2$), index $\rm ind$=2, and cell = END.
  \While 1
  \State $\mathcal{T}^*(\rm ind -1)$ = cell
  \State cell= $P_{\dist}(\rm ind-1, st-ind +1)$
  \If {cell = START}
     \State $\mathcal{T}^*(\rm ind)$ = START 
     \State Break
  \EndIf
  \State Set ind=ind+1.
  \EndWhile
  \State Return the optimal path stored in $\mathcal{T}^*$, representing the array of cells that the cargo-UAV should visit to reach the END cell.
\end{algorithmic}
\end{algorithm}

\begin{figure}[t]
\centering
\includegraphics[width=0.6\linewidth]{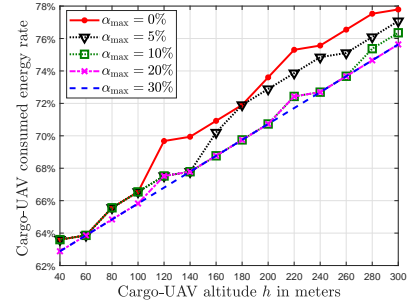}
\caption{Consumed energy rate vs. $h$ (different $\alpha_{\max}$).}
\label{fig:bat}
\end{figure}

Fig. \ref{fig:bat} shows the consumed energy rate, defined as the ratio of the consumed energy for the mission to the available energy $E_C-E_S$, for the setup of Fig. \ref{fig:path}.
This performance metric is evaluated as a function of the flying altitude $h$ and for different disconnectivity rates $\alpha_{\max}$.
As $\alpha_{\max}$ becomes more stringent, the cargo-UAV consumes more energy in order to execute detours and avoid coverage holes, thus agreeing with the results of Fig. \ref{fig:path}. 
Also, the energy consumption increases with the altitude. Indeed, in addition to the energy consumed between START and END locations, the cargo-UAV has to spend more effort in takeoff and landing to reach them.

\begin{figure}[t]
\centering
\includegraphics[width=0.55\linewidth]{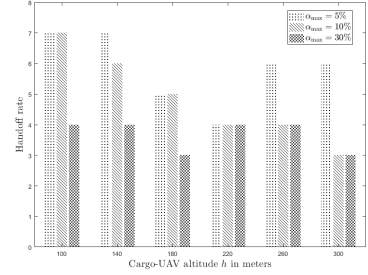}
\caption{Handoff rate vs. $h$ (different $\alpha_{\max}$).}
\label{fig:ho_height}
\end{figure}

In Fig. \ref{fig:ho_height}, we present the handoff rate as a function of $h$ and for different $\alpha_{\max}$, given the setup of Fig. \ref{fig:path}. We notice no clear trend or correlation can be identified between the handoff rate and the cargo-UAV altitude. However, the handoff rate increases as $\alpha_{\max}$ decreases, given a fixed $h$. Indeed, as $\alpha_{\max}$ nears zero, i.e., no disconnectivity is allowed, the cargo-UAV is required to follow paths that avoid coverage holes, i.e., it is more likely to trigger handoff events when travelling.

\begin{figure}[t]
\centering
\includegraphics[width=0.56\linewidth]{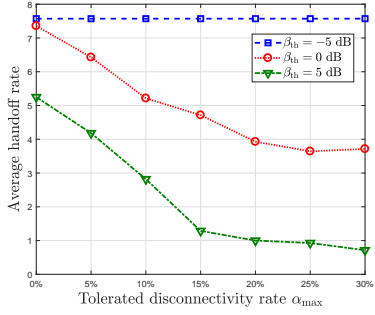}
\caption{Avg. handoff rate vs. $\alpha_{\max}$ (different $\beta_{\rm th}$).}
\label{fig:hodisc}
\end{figure}

Fig. \ref{fig:hodisc} illustrates the average handoff rate (over different cargo-UAV altitudes) as a function of $\alpha_{\max}$, and for different \textsf{SIR} threshold values $\beta_{\rm th}$. As $\alpha_{\max}$ increases, the handoff rate improves. Indeed, with $\alpha_{\max}$ loosened, the cargo-UAV can cross a higher number of coverage holes, which does not account for handoff events. In contrast, a stringent $\alpha_{\max}$ obligates the cargo-UAV to move only within the BSs' coverage areas, thus increasing the handoff rate.
Moreover, the handoff rate degrades as $\beta_{\rm th}$ decreases. In fact, a small $\beta_{\rm th}$ means that a small number of (or no) coverage holes are present within the network. Thus, the cargo-UAV experiences frequent handoffs when travelling. 
Finally, we notice that the average handoff rate is independent from $\alpha_{\max}$ when $\beta_{\rm th}=-5$ dB. Indeed, for this $\beta_{\rm th}$ value, no coverage holes are present, and hence no disconnectivity is encountered in the cargo-UAV path.

\begin{figure}[t]
\centering
\includegraphics[width=0.56\linewidth]{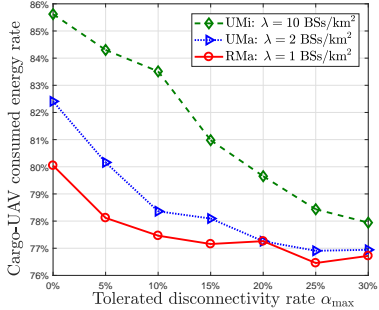}
\caption{Avg. consumed energy rate vs. $\alpha_{\max}$ ($h=100$ m, different environments).}
\label{fig:env}
\end{figure}

We investigate in Fig. \ref{fig:env} the average consumed energy rate as a function of $\alpha_{\max}$, for different types of environments, namely UMa, UMi, and RMa, characterized by the BSs density $\lambda$, and where the results are averaged over 200 Monte-Carlo realisations of BSs locations.
First, the energy performance improves with $\alpha_{\max}$ in any environment type. This is expected since a more direct path is followed by the cargo-UAV for higher $\alpha_{\max}$. Second, RMa achieves the best consumed energy rate for any $\alpha_{\max}$. Indeed, in RMa, a small number of BSs is used, which limits inter-cell interference and hence improves the coverage of the BSs. Consequently, short paths are followed by the cargo-UAV to reach its destination. However, as the environment becomes denser with BSs, the impact of inter-cell interference amplifies due to LoS links, causing a significant degradation to communication channels and coverage areas. Hence, the cargo-UAV has to make detours in order to reach its destination with respect to $\alpha_{\max}$.

\section{Conclusion}
In this paper, we investigated the disconnectivity and handoff aware path planning problem for the cargo-UAV, aiming to minimize both its energy consumption and handoff rate.
The formulated complex problem is solved using a dynamic programming approach, which achieves optimal flying trajectories. 
Through simulations, we establish the relationship between cellular disconnectivity and handoff rate for aerial communications. 
Indeed, a more stringent disconnectivity requirement increases both the handoff rate and the energy consumption due to longer trajectories. Also, the impact study of different parameters provides the following guidelines: 1) Given a high tolerable disconnectivity rate, it is preferable to fly at the lowest permissible altitude in the most direct path to reduce energy consumption, 2) to operate partially or fully independently from the disconnectivity rate, it is encouraged to use a high-sensitive receiver at the cargo-UAV (which corresponds to a low $\beta_{\rm th}$), however a high handoff rate is expected, and
3) for energy-efficient operation, it is recommended to operate in low-density areas, e.g., RMa, or to redesign the frequency strategy in high-density areas, e.g., UMi, in order to reduce inter-cell interference.

\bibliographystyle{IEEEtran}  
\bibliography{references,Bibliography}

\end{document}